\documentclass[print]{revtex4}
\usepackage{amsmath,amssymb}
\usepackage{graphicx}

\begin{document}

\title{\large \bf  Optical generation of hybrid entangled state via entangling single-photon-added coherent state}

\author{Yan Li,$^{1,2,3}$\footnote{Electronic address: liyan@wipm.ac.cn} Hui Jing,$^{1,2}$ and Ming-Sheng Zhan$^{1,2}\footnote{Electronic address: mszhan@wipm.ac.cn}$}

\affiliation{$^{1}$State Key Laboratory of Magnetic Resonance and
Atomic and Molecular Physics,\\
 Wuhan Institute of Physics and Mathematics, Chinese Academy of Sciences, Wuhan 430071, P. R. China\\
 $^{2}$Center for Cold Atom Physics, Chinese Academy of Sciences, Wuhan
 430071, P. R. China\\
$^{3}$Graduate School of the Chinese Academy of Sciences, Beijing
100080, P. R. China}

\begin{abstract}
We propose a feasible scheme to realize the optical entanglement
of single-photon-added coherent state (SPACS) and show that,
besides the Sanders entangled coherent state, the entangled SPACS
also leads to new forms of hybrid entanglement of quantum Fock
state and classical coherent state. We probe the essential
difference of two types of hybrid entangled state (HES). This HES
provides a novel link between the discrete- and the
continuous-variable entanglement in a natural way.\\

PACS numbers: 03.67.Mn 42.50.Dv 03.65.Ud
\end{abstract}

\baselineskip=16pt

\maketitle

The generation of quantum entangled states plays an important role
in quantum information science \cite{1}, and many practical
schemes have been investigated theoretically and experimentally by
applying, e.g., light fields \cite{2}, trapped ions \cite{3},
cavity QED \cite{4} and ultra-cold atomic ensembles \cite{5}, etc.
Among these appealing schemes, the optical generations of
entanglement are always of intense interests due to their
versatile applications for various purposes \cite{6}, including
the well-known discrete-variable \cite{7,8,9,10,11} and the
so-called continuous-variable entangled states
\cite{12,13,14,15,16,17}. Therefore a simple question may
naturally arise: is it possible (and then useful) to optically
generate an $intermediate$ hybrid entanglement which, however, may
not be reduced to any of these two types of entangled states?

Recently, Zavatta \emph{$et~al.$} in their beautiful experiment
(Science 306 (2004) 660) generate a novel single-photon-added
coherent state (SPACS) and then visualize the interesting
evolution of quantum-to-classical transition \cite{18}. The key
elements of their experiment are the BBO-crystal-based parametric
down-conversion process and then the single-photon detection
technique \cite{18}. This experiment is also a clear
implementation of the original idea of Agarwal \emph{$et~al.$
}\cite{19} of the photon-added coherent state (PACS) via
successive elementary one-photon excitation of a classical
coherent field, for example \cite{18},
\begin{eqnarray}\label{eqn:1}
SPACS:~~|\alpha,1\rangle=\frac{\hat{a}^{\dag}|\alpha\rangle}{\sqrt{1+|\alpha|^2}},
\end{eqnarray}
where $|\alpha\rangle$ is the ordinary Glauber coherent state and
$\hat{a}^{\dag}$ is the photon creation operator. These new states
occupy an intermediate position between the fully
quantum-mechanical single-photon Fock state and the classical
coherent state, containing both the discrete and continuous
variables in some sense.

 In this paper, we propose a feasible scheme to optically realize the
 entangled SPACS (ESPACS) and show that it can lead to some hybrid
 entanglement of single-photon Fock state and classical coherent state.
Our scheme is based on the combination of two elegant concepts:
the SPACS and the well-known entangled coherent state (ECS)
firstly proposed by Sanders \cite{20}.  We will show that, by
generating the ESPACS in the light field, the interesting
entanglement between the discrete and continuous variables can be
achieved, which is the so called hybrid entangled state (HES). We
note that very recently Feng $et~al.$ suggested to generate a
special entanglement of the parity qubit (even or odd coherent
states) and the spin qubit (discrete variables) via the technique
of conditional joint measurement \cite{21}. As an interesting
comparison, the output HES in our scheme has some very different
characteristics, e.g., the qubit itself is $hybrid$ in the sense
that one cannot clarify which is the spin or the parity qubit in
the created entanglement. In addition, our scheme is feasible
which only needs coherent input lights and the single-photon
detections. Of course, the unattractive point is the use of some
nonlinear mediums.

Turning now to Fig. $1$ for an illustration of our schematic
setup.
\begin{figure}[ht]
\includegraphics[width=0.8\columnwidth]{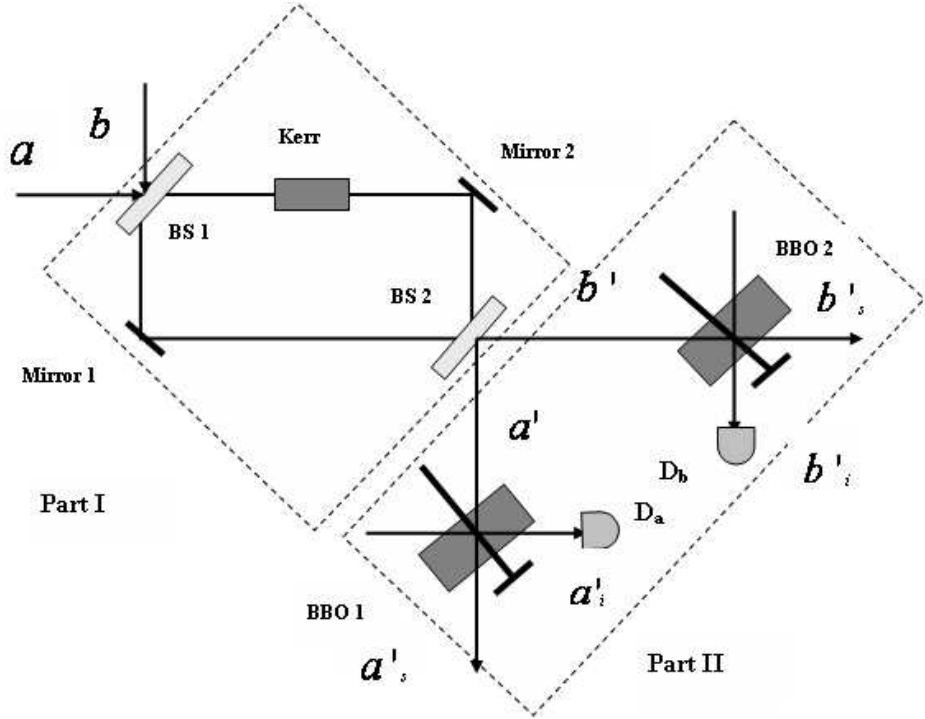}
\caption{Configuration of the generation of the entangled
single-photon-added coherent state. Part I is composed of the two
beam splitters, one Kerr medium and two mirrors, while part II
contains two BBO crystals and two single-photon detectors.}
\end{figure}
 We can divide this configuration into two parts: the
Mach-Zehnder (MZ) interferometer (I) and the BBO crystals (II).
Let us firstly consider the part I. Two beams of classical
coherent fields are incident on the first beam splitter
($BS_{1}$). Within one arm of the interferometer a nonlinear Kerr
medium is placed which we approximate as a nonlinear oscillator in
a single-mode treatment. For simplicity we assume that the
nonlinear interferometer is lossless. The dynamical description
involves two input modes $a$ and $b$, with corresponding Bose
annihilation operators $\hat{a}$ and $\hat{b}$. This indicates the
two-mode initial state as the following
\begin{eqnarray}\label{eqn:2}
|\psi\rangle_{in1}=|\alpha\rangle_a|\beta\rangle_b.
\end{eqnarray}
 The BS Hamiltonian generating linear mode-coupling is given by
\cite{22}
\begin{eqnarray}\label{eqn:3}
\hat{H}_{B}=i(\lambda^*\hat{a}^\dag\hat{b}-\lambda\hat{a}\hat{b}^\dag),
\end{eqnarray}
where $\lambda$ denotes the coupling strength between the two
modes and $arg(\lambda)$ denotes the relative phase shift between
the modes imposed by the coupling. The unitary evolution operator
is
\begin{eqnarray}\label{eqn:4}
\hat{U}_{B}(\theta,\phi)=\exp[\theta(e^{-i\phi}\hat{a}^\dag\hat{b}-e^{i\phi}\hat{a}\hat{b}^\dag)],
\end{eqnarray}
where $\theta$=$\lambda$$t$, with  $t$ being the interaction time
and $\phi$=$arg(\lambda)$. This unitary operator keeps the
factorized structure of the state of the system by transforming a
two-mode product coherent state (CS) into another CS, which means
that the state after the first BS is
\begin{eqnarray}\label{eqn:5}
\hat{U}_B(\theta,\phi)|\psi\rangle_{in1}=|\alpha\cos\theta+\beta{e^{-i\phi}}\sin\theta\rangle_a|-\alpha{e^{i\phi}}\sin\theta+\beta\cos\theta\rangle_b.
\end{eqnarray}
For the case of $\theta$=$\pi$/$4$ and $\phi$=$3$$\pi$/$2$, we get
\begin{eqnarray}\label{eqn:6}
\hat{U}_B(\theta,\phi)|\psi\rangle_{in1}=|{\frac{1}{\sqrt{2}}}(\alpha+i\beta)\rangle_a|{\frac{1}{\sqrt{2}}}(\beta+i\alpha\rangle)_b.
\end{eqnarray}
\indent The Hamiltonian of nonlinear Kerr medium is $
\hat{H}_{K}=\chi(\hat{a}^\dag\hat{a})^2$, the unitary evolution
operator is
\begin{eqnarray}\label{eqn:8}
\hat{U}_{K}=\exp[-i\chi(\hat{a}^\dag\hat{a})^2],
\end{eqnarray}
where $\chi$ is the nonlinearity coefficient which is proportional
to the nonlinear coefficient $\chi^{(3)}$ of the medium and the
interaction length. For a CS $|\alpha\rangle_{a}$, the state after
the Kerr interaction is ($\chi=\pi/2$)
\begin{eqnarray}\label{eqn:9}
\hat{U}_{K}|\alpha\rangle_a=
\exp[-|\alpha|^2/2]{{\sum^{\infty}_{n=0}}\frac{\exp[-i\frac{\pi}{2}n^2]\alpha^n|n\rangle}{\sqrt{n!}}}
=\frac{1}{\sqrt{2}}(e^{-i\pi/4}|\alpha\rangle+e^{i\pi/4}|-\alpha\rangle).
\end{eqnarray}
\indent Combining the Kerr medium and two BS in the MZ
interferometer, we obtain the Sanders ECS \cite{20} as the output
state of part I:
\begin{eqnarray}\label{eqn:10}
|\psi\rangle_{out1}=\hat{U}_{MZ}|\psi\rangle_{in1}=
\hat{U}_B\hat{U}_K\hat{U}_B|\alpha\rangle_a|\beta\rangle_b=
\frac{1}{\sqrt{2}}(e^{-i\pi/4}|i\beta\rangle_{a'}|i\alpha\rangle_{b'}+e^{i\pi/4}|-\alpha\rangle_{a'}|\beta\rangle_{b'}).
\end{eqnarray}

In our scheme the Sanders ECS $|\psi\rangle_{out1}$ is used as the
$input$ state of Part II in which two nonlinear crystals provide
the important further manipulations. For the parametric
down-conversion of BBO crystal, one high-energy pump photon can
induce two lower-energy photons in symmetrically oriented
directions being called the signal and idler modes. Without other
light being injected into the crystal, a pair of entangled photons
with random but correlated phases is produced. The Hamiltonian of
BBO crystal is $
\hat{H}_{C}=\kappa\hat{c}^{\dag}\hat{a}\hat{b}+\kappa\hat{a}^{\dag}\hat{b}^{\dag}\hat{c},
$ where the operator $\hat{c}$ can be regarded as $c$-number
${i\gamma}$ for strong (classical) pumping, thereby the
corresponding evolution operator is
\begin{eqnarray}\label{eqn:12}
\hat{U}_c=\exp[-i(\kappa\hat{c}^{\dag}\hat{a}\hat{b}+\kappa\hat{a}^{\dag}\hat{b}^{\dag}\hat{c})t]=\exp[{{\kappa\gamma}t}(\hat{a}^{\dag}\hat{b}^{\dag}-\hat{a}\hat{b})]=\exp[g(\hat{a}^{\dag}\hat{b}^{\dag}-\hat{a}\hat{b})],
\end{eqnarray}
where $g$=${\kappa\gamma}t$ can be regarded as an effective
interaction time. The two-mode entanglement is obtained for the
input CS signal and vacuum idler states (see Fig. $2$).
\begin{figure}[ht]
\includegraphics[width=0.45\columnwidth]{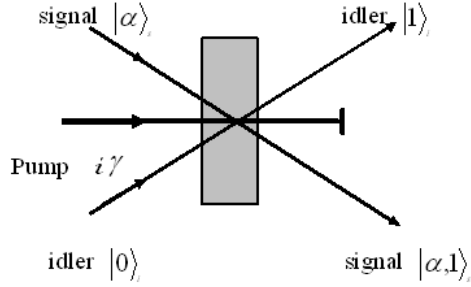}
\caption{The generation of entangled photon-pairs in the oriented
directions via the parametric down-conversion of pump light
$i\gamma$. Given coherent state input in one mode and the other
vacuum state, when one photon is detected in the idler mode, the
SPACS $|\alpha,1\rangle$ will be created in the signal mode.}
\end{figure}
If the parametric gain is kept sufficiently low, i.e. $|g|\ll1$,
the form of output state can be written as
\begin{eqnarray}\label{eqn:14}
\hat{U}_{c}|\alpha\rangle_{s}|0\rangle_{i}=
\exp[g(\hat{a}^{\dag}\hat{b}^{\dag}-\hat{a}\hat{b})]|\alpha\rangle_{s}|0\rangle_{i}
\approx|\alpha\rangle_{s}|0\rangle_{i}+g\hat{a}^{\dag}|\alpha\rangle_{s}|0\rangle_{i}=
|\alpha\rangle_{s}|0\rangle_{i}+g|\alpha,1\rangle_{s}|1\rangle_{i}.
\end{eqnarray}

The output signal mode will contain the original CS except for the
relatively rare single-photon detections in the idler output mode.
These rare events stimulate emission of one photon in the CS
$|\alpha\rangle$ \cite{18}, which generates the intermediate state
$|\alpha,1\rangle_{s}$ in the correlated signal mode. Therefore,
considering the second part of the device shown in Fig. $1$, the
input state for part II is: $
|\psi\rangle_{in2}=|\psi\rangle_{out1}
=\hat{U}_{MZ}|\alpha\rangle_{a}|\beta\rangle_{b}$, and the final
output state after the BBO interactions for the two modes $a'$ and
$b'$ can be obtained as
$$|\psi\rangle_{out}
=\hat{U}_{ca'}\hat{U}_{cb'}\hat{U}_{MZ}|\alpha\rangle_{a}|\beta\rangle_{b}\\
=\exp[g_{1}(\hat{a}^{\dag}_{s}\hat{a}^{\dag}_{i}-
\hat{a}_{s}\hat{a}_{i})]\exp[g_{2}(\hat{b}^{\dag}_{s}\hat{b}^{\dag}_{i}-
\hat{b}_{s}\hat{b}_{i})]\hat{U}_{MZ}|\alpha\rangle_{a}|\beta\rangle_{b}\\$$
\begin{eqnarray}\label{eqn:16}
=(1+g_{1}\hat{a}^{\dag}_{s}\hat{a}^{\dag}_{i})
(1+g_{2}\hat{b}^{\dag}_{s}\hat{b}^{\dag}_{i})\frac{1}{\sqrt{2}}
(e^{-i\pi/4}|i\beta\rangle_{a's}|0\rangle_{a'i}|i\alpha\rangle_{b's}|0\rangle_{b'i}+e^{i\pi/4}|-
\alpha\rangle_{a's}|0\rangle_{a'i}|\beta\rangle_{b's}|0\rangle_{b'i}).
\end{eqnarray}

For our present purpose we can simply assume that the whole device
is lossless and let the effective interaction time of two BBO
crystals just be equal, i.e. $g_{1}=g_{2}$, then we reach the
following output entangled state in a rather general $hybrid$
form:
\begin{eqnarray}
|\psi\rangle_{out}&=&\frac{1}{\sqrt{2}}
[e^{-i\pi/4}|i\beta\rangle_{a's}|0\rangle_{a'i}|i\alpha\rangle_{b's}|0\rangle_{b'i}+
e^{i\pi/4}|-\alpha\rangle_{a's}|0\rangle_{a'i}|\beta\rangle_{b's}|0\rangle_{b'i}]\nonumber\\
&&+\frac{1}{\sqrt{2}}g[e^{-i\pi/4}|i\beta,1\rangle_{a's}|1\rangle_{a'i}|i\alpha\rangle_{b's}
|0\rangle_{b'i}+e^{i\pi/4}|-\alpha\rangle_{a's}|0\rangle_{a'i}|\beta\rangle_{b's}|0\rangle_{b'i}]\nonumber\\
&&+\frac{1}{\sqrt{2}}g[e^{-i\pi/4}|i\beta\rangle_{a's}|0\rangle_{a'i}|i\alpha,1\rangle_{b's}|1\rangle_{b'i}+
e^{i\pi/4}|-\alpha\rangle_{a's}|0\rangle_{a'i}|\beta,1\rangle_{b's}|1\rangle_{b'i}]\nonumber\\
&&+\frac{1}{\sqrt{2}}g^{2}[e^{-i\pi/4}|i\beta,1\rangle_{a's}|1\rangle_{a'i}|i\alpha,1\rangle_{b's}
|1\rangle_{b'i}+e^{i\pi/4}|-\alpha,1\rangle_{a's}|1\rangle_{a'i}|\beta,1\rangle_{b's}|1\rangle_{b'i}].
\end{eqnarray}
This indicates the conditional preparations of the different kinds
of entangled single-photon-added CS (ESPACS) whenever a "click" is
registered or not on the single-photon detectors placed in the
output idler modes. For the concrete illustrations, we start to
analyze in the following four different circumstances of this output
entangled states:\\
\indent First, if both of the two detectors do not detect one
photon synchronously, we just get the entangled coherent state
(ECS) of the output signal modes \cite{20}, as it should be:
\begin{equation}
|\psi\rangle_{out}=\frac{1}{\sqrt{2}}[e^{-i\pi/4}|i\beta\rangle_{a's}|i\alpha\rangle_{b's}
+e^{i\pi/4}|-\alpha\rangle_{a's}|\beta\rangle_{b's}].
\end{equation}

Second, if one photon is registered in one detector (detector $a$
or $b$ as case $A$ or $B$) but not in the other one, we can get
the generalized ECS, i.e., the entanglement between the SPACS and
the CS:
\begin{equation}
|\psi\rangle_{out}=\left \{
\begin{array}{ll}
\frac{1}{\sqrt{2}}g[e^{-i\pi/4}|i\beta,
1\rangle_{a's}|i\alpha\rangle_{b's}+e^{i\pi/4}|-\alpha,1\rangle_{a's}|\beta\rangle_{b's}]:
& {\rm case~ A;\;} \\
\frac{1}{\sqrt{2}}g[e^{-i\pi/4}|i\beta\rangle_{a's}|i\alpha,1\rangle_{b's}+
e^{i\pi/4}|-\alpha\rangle_{a's}|\beta,1\rangle_{b's}] : & {\rm
case~ B, }
\end{array}
\right.
\end{equation}
which can be termed as the type-II hybrid entangled state (HES),
since it is essentially different from the type-I HES obtained by
Feng $et~al.$ (i.e., the entanglement of the parity qubit and the
spin qubit) via their BS-based conditional joint measurement way
\cite{21} or Chen $et~al.$ via non-optics methods \cite{23}.

Finally, if both of the detectors can capture one photon
simultaneously, we can obtain the interesting entanglement between
the intermediate single-photon-added CS (ESPACS), i.e.,
\begin{equation}
ESPACS:~~\frac{1}{\sqrt{2}}g^{2}[e^{-i\pi/4}|i\beta,1\rangle_{a's}|i\alpha,1\rangle_{b's}
+e^{i\pi/4}|-\alpha,1\rangle_{a's}|\beta,1\rangle_{b's}].
\end{equation}
This state, as a generalized form of ECS, realizes the
entanglement of SPACS in a conceptually elegant way. In
particular, if we let $\beta=0$ and assume $\alpha$ being large
enough, or equivalently the initial input state is simple:
$|\psi\rangle_{in1}=|\alpha\rangle_{a}|0\rangle_{b}$, we can see
that, by choosing suitable parameters, the simplified form of
ESPACS leads to the HES between discrete and continuous variables:
\begin{equation}
|\Psi^\omega_{a,b}\rangle_{HES-II}=
|1\rangle_{a}|\alpha\rangle_{b}+e^{i\omega}|\alpha\rangle_{a}|\tilde{1}\rangle_{b},
\end{equation}
here $\omega$ is the corresponding phase factor, and the tilde
denotes "$1\rightarrow 0$" for the second case of the above three
examples. Obviously, this denotes a very interesting $new$ form of
HES since it cannot be logically encoded into any type of
discrete-variable entangled state like the type-I HES \cite{21,
23}, thus it can be taken as the type-II HES or the entanglement
of the $hybrid~ qubits$ which, in our opinions, is the essential
hybrid entanglement of discrete variables (DVs) and continuous
variables (CVs).

We note that in two recent related works, Chen \emph{et al.}
probed the hybrid entangled state (HES) in the trapped-ion and the
atom-cavity systems \cite{23}, and Feng \emph{et al.} investigated
the mixed entangled state (MES) via an optical scheme \cite{21},
these two states are in fact the same form, i.e.,
\begin{equation}
|\Psi^{\pm}_{1,2}\rangle_{HES-I}=
\xi(|\uparrow\rangle_{1}|\alpha\rangle_{o,2}\pm|\downarrow\rangle_{1}|\alpha\rangle_{e,2}),
\end{equation}
where $\xi$ is a normalized factor and $|\alpha\rangle_{o,e}$
denotes the odd or even CS. Clearly, this is just the formally
discrete-variable entanglement of spin qubit and parity qubit. In
other words, the physically CVs are logically encoded into the
formally DVs or parity qubit \cite{21, 23}. However, in our
type-II HES, we cannot tell which is the spin or the parity qubit.
The qubits itself are hybrid or mixed. In some sense, we can view
the type-II HES as the entangled Schr$\ddot{o}$dinger's cat state
if we define the classical-world CVs and the quantum-world DVs as
the living-cat or the dead-cat states, respectively.

In conclusion, we propose a feasible scheme to achieve optical
entanglement of SPACS and thereby in a conceptually elegant way
show that, besides the original Sanders ECS as the CV entanglement
of two CS, the generated ESPACS also leads to some new forms of
HES of purely quantum Fock state and classical CS. Our scheme is
quite simple by combining two famous and experimentally accessible
techniques. It consists only of three kinds of familiar devices:
two BSs -- providing two-mode input-output ports, nonlinear Kerr
medium -- generating CV entanglement of CS, and BBO crystals --
providing DV entanglement of photon pairs. We compared our new
type of HES (type II HES) with two previous related states (type-I
HES) and pointed out their essential differences, i.e., whether or
not they can be written as the form of spin- and parity-qubit
entanglement \cite{24}. These new forms of HES can also be
expected to be realized in other non-optical systems. Besides
providing a natural link between the DV and the CV entanglements,
these HES may serve as new entanglement resources and their novel
applications in quantum information science would be challenging
for further studies.

\bigskip

\noindent We gratefully acknowledge helpful discussions with Pof.
Hongwei Xiong and Dr. Min Liu. This work was supported by NFSC
(Grant Nos. 10304020, 10474119 and 10474117), National Basic
Research Programme of China (Grant No. 2001CB309309) and the fund
from Chinese Academy of Sciences.

%\section{}

%\section{Results}
%\section{Conclusions}
%\indent

%%%%%%%%%%%%%%%%%%%%%%%%%%%%%%%%%%%%%%%%%%%%%

%%%%%%%%%%%%%%%%%%%%%%%%%%%%%%%%%%%%%%%%%%%%%

%\bigskip

\end{document}